\documentclass[conference]{IEEEtran}
\IEEEoverridecommandlockouts
\usepackage{cite}
\usepackage{pifont}
\usepackage{setspace}
\newcommand{\cmark}{\ding{51}} 
\newcommand{\xmark}{\ding{55}} 
\usepackage{threeparttable}
\usepackage{amsmath,amssymb,amsfonts}
\usepackage[linesnumbered,ruled,vlined]{algorithm2e}
\usepackage{algorithmic}
\usepackage{listings}
\usepackage{graphicx}
\usepackage{textcomp}
\usepackage{tabularx}
\usepackage{booktabs}
\usepackage{tcolorbox}
\usepackage{xcolor}
\def\BibTeX{{\rm B\kern-.05em{\sc i\kern-.025em b}\kern-.08em
    T\kern-.1667em\lower.7ex\hbox{E}\kern-.125emX}}
\begin{document}

\title{LLM-Based Semantic Modeling and Cooperative Evolutionary Fuzzing for Traffic Violation Scenario Generation}

\author{
Yangyang Liu$^{1,2}$, Xinyu Li$^{1,2}$, Yan Xiao$^{3, *}$, Miao Zhang$^{4}$, Pengcheng Zhang$^{1,2}$\\
\thanks{*Corresponding author: Yan Xiao (email: xiaoy367@mail.sysu.edu.cn)}\\
$^{1}$Key Laboratory of Water Big Data Technology of Ministry of Water Resources, Hohai University, Nanjing, China\\
$^{2}$College of Computer Science and Software Engineering, Hohai University, Nanjing, China\\
$^{3}$School of Cyber Science and Technology, Sun Yat-sen University, Shenzhen, China\\
$^{4}$School of Artificial Intelligence, Shenzhen Technology University, Shenzhen, China\\
\text{\{youngyliu, lixiny, pchzhang\}@hhu.edu.cn, xiaoy367@mail.sysu.edu.cn, zhangmiao@sztu.edu.cn}
}

\maketitle

\begin{abstract}
Ensuring the safety of autonomous driving systems (ADS) in a cost-effective and efficient manner remains a critical challenge. Existing law-guided scenario generation approaches are typically limited to a narrow subset of legal rules, resulting in insufficient scenario diversity, and search-based methods often struggle with large and sparse search spaces. To address these limitations, we propose SLaFE (\underline{S}emantic \underline{La}w Modeling and \underline{F}uzzing based on Cooperative \underline{E}volution), a novel traffic violation scenario generation framework designed to systematically evaluate the safety of ADS.
SLaFE harnesses the reasoning capabilities of large language models (LLMs) to convert traffic laws into structured scenario constraints. These scenarios are then optimized via a cooperative evolutionary fuzzing algorithm that explores the parameter space to identify boundary cases likely to trigger abnormal ADS behaviors. We evaluate SLaFE on the Apollo platform within the LGSVL simulator using ten real-world traffic regulations. 
Experimental results show that SLaFE successfully triggered all 10 types of traffic law violations (10/10), outperforming the best existing method, VioHawk (9/10), while others detected no more than 3. Moreover, SLaFE achieved an average triggering time of 5.1 minutes per law type, significantly faster than VioHawk (9.0 minutes) and other baselines. These results highlight SLaFE’s effectiveness in discovering diverse and critical law-violating scenarios for ADS testing.
\end{abstract}


\section{Introduction}

Ensuring the safety and reliability of ADS requires rigorous testing to uncover potential risks before deployment~\cite{li2023simulation,baresi2024efficient,woodlief2024s3c}. Although real-world on-road testing provides valuable insights, it is inherently expensive, time-consuming, and often lacks sufficient coverage of rare or safety-critical scenarios. As a scalable alternative, simulation-based testing enables ADS evaluation under a wide spectrum of traffic conditions—ranging from routine situations to rare, safety-critical corner cases~\cite{zhang2023building,cheng2024decictor}. Recent research emphasizes the importance of constructing realistic, diverse, and semantically meaningful scenarios to better reflect the complexities of real-world driving~\cite{cheng2023behavexplor,haq2023many,haq2022efficient}.

The effectiveness of simulation-based testing largely depends on the quality and semantic relevance of the generated scenarios~\cite{woodlief2025differential,hildebrandt2023physcov}. Traditional methods, including manual design and collision-based scenario generation~\cite{kim2022drivefuzz,tian2022mosat,tang2024legend}, often lack a principled integration in real-world traffic semantics, limiting their ability to produce meaningful edge-case scenarios. In contrast, traffic laws define normative constraints such as yielding obligations, right-of-way rules, and stopping requirements that govern legal and illegal driving behaviors~\cite{sun2022lawbreaker}. These laws, although expressed in natural language, provide a rich semantic foundation for generating violation-triggering test cases with high realism and coverage.

Traffic regulations are typically written in natural language, exhibiting inconsistency in expression and ambiguity in semantics, which makes them difficult to directly apply in simulation-based scenario generation. This necessitates a specification framework capable of adaptively transforming diverse legal texts into machine-interpretable representations. To this end, prior studies such as Lawbreaker~\cite{sun2022lawbreaker} and ABLE~\cite{zhang2023testing} have introduced domain-specific languages (DSLs) to formally model traffic laws in a standardized form. However, these approaches are often tightly coupled with the legal texts, requiring developers to manually reinterpret and re-encode constraints whenever new scenarios or regulations are introduced. Such semantic entanglement significantly limits the scalability and applicability of these frameworks.

Beyond linguistic ambiguity, the heterogeneity of traffic laws across jurisdictions further complicates their formalization. Such variations manifest not only in structural forms but also in legal interpretations across different countries and regions. When formalized, this diversity translates into distinct optimization objectives, resulting in challenges such as ambiguous search targets and redundant computations during scenario generation. While some methods attempt to mitigate these issues by modeling formalized rules as independent units~\cite{sun2022lawbreaker,li2024viohawk}, or by unifying them into a single evaluation metric~\cite{zhang2023testing}, they often fail to recognize and exploit the structural and behavioral commonalities shared across different regulations. As a result, these methods suffer from poor knowledge reuse, fragmented optimization goals, and ultimately, significant efficiency bottlenecks.

To address these challenges, we propose SLaFE, a novel testing approach for ADS that integrates LLM-guided semantic law abstraction and cooperative evolutionary fuzzing for scenario generation. SLaFE employs Violation Modeling Based on Regional Semantics as its specification framework and adopts a two-stage prompting mechanism to model traffic laws. In the first stage, an LLM is used to infer key situational elements from traffic laws, forming a coarse-grained scenario description. In the second stage, this description guides the LLM to generate violation feature vectors (VFVs), which are then used to construct initial scenario instances. To enhance generation efficiency and legal adaptability, we design a cooperative evolutionary algorithm that categorizes traffic laws with similar regional semantics. A fuzzing algorithm is then applied to iteratively optimize the initial scenarios, generating high-risk violation test cases for evaluating the robustness of ADS. We evaluate our method on the Apollo~\cite{Baidu2019} and LGSVL~\cite{rong2020lgsvl} simulation platforms. The results demonstrate that our approach effectively generates semantically meaningful and behaviorally diverse violation scenarios. Compared to the average performance of baseline methods, it achieves approximately a 62\% improvement in violation type coverage and a 54\% increase in violation-triggering efficiency. Our main contributions are as follows:

\begin{itemize}
    \item We propose a traffic rule transformation framework that leverages LLMs to support the safety validation of ADS under complex legal and physical driving environments.
    \item We introduce a regulation categorizing strategy based on regional semantics, which significantly improves test generation efficiency compared to existing approaches.
    \item We infer violation spaces from traffic regulations to guide the direction of scenario evolution, effectively narrowing the search space and reducing unnecessary computational overhead. Our codes are open-sourced\footnote{https://github.com/YangyangLiu-Research/SLaFE}.
\end{itemize}

\section{Background And Motivation}

\subsection{Prompt Engineering}

LLMs have emerged as effective tools for information extraction in domain-specific tasks, offering the capability to transform unstructured legal text into structured representations. This opens a new direction for testing ADS in terms of their ability to violate laws. We leverage domain-specific prompts tailored to autonomous driving to guide the LLM in extracting key scenario-relevant elements from legal texts, such as their spatial configurations, relative positions, speeds, and intended behaviors. The extracted information is then encoded into structured scenario vectors, which serve as the basis for initializing simulation environments. Building upon these initial scenarios, our custom-designed fuzzing algorithm iteratively modifies scenario parameters to generate challenging test cases.

\subsection{Motivation}

In ADS testing, ensuring compliance with traffic laws, beyond simply avoiding collisions, has become a crucial safety objective~\cite{sun2022lawbreaker}. Unlike the straightforward objective of collision avoidance, law compliance involves diverse semantics and region-specific interpretations. Consequently,  existing law-based testing methods often require dedicated oracles for individual rules, resulting in fragmented workflows and turning fuzzing into a complex multi-objective search problem.

Recent methods such as VioHawk~\cite{li2024viohawk} address this by introducing region-based testing, which maps each regulation to high-risk zones where violations are likely to occur. By narrowing the admissible driving space and steering the ADS toward these zones, region-based testing simplifies rule modeling and significantly enhances the efficiency of violation discovery.

However, region-based testing still models and tests each regulation independently, limiting knowledge reuse. We observe that under similar road structures, different laws often share semantically similar or even identical high-risk regions. This suggests that a region-semantic modeling approach could unify regulation-specific tests and enable collaborative violation discovery across laws.

\begin{figure}[htbp]
    \centering
    \includegraphics[width=0.41\textwidth]{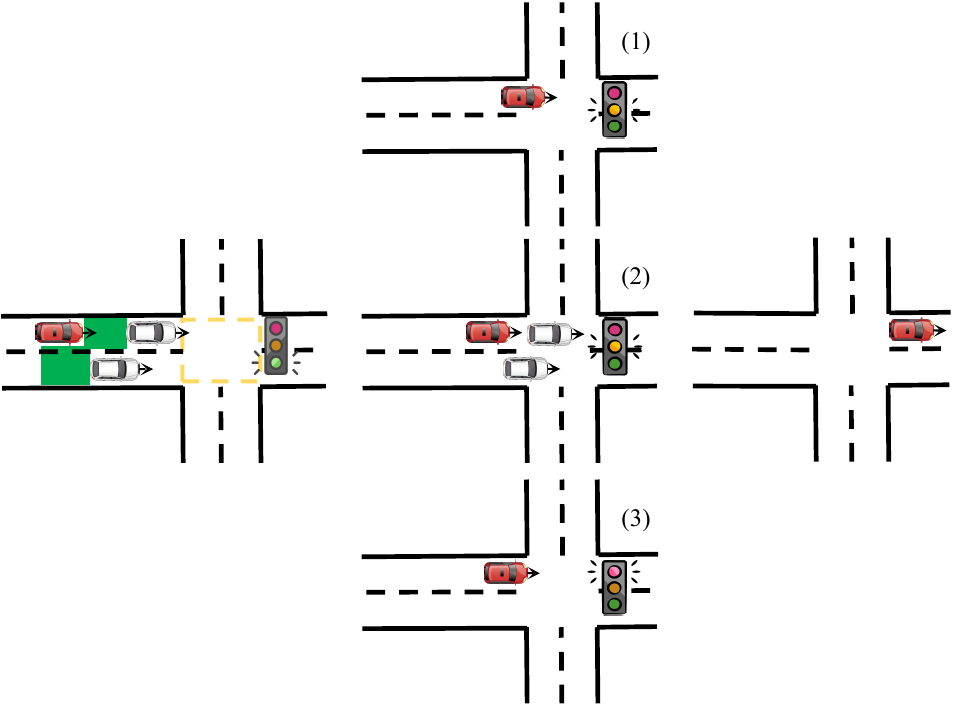}  
    \caption{Shared region semantics across different regulations}
    \label{fig:motivation}
\end{figure}

Fig.~\ref{fig:motivation} illustrates this phenomenon in an intersection scenario, where the ego vehicle is guided to stop before the stop line. Depending on the traffic light status, three distinct legal interpretations arise: (1) the vehicle has already crossed the line when the yellow light turns on and proceeds legally; (2) the yellow light turns on but the vehicle has not yet crossed, and continues moving—violating the yellow-light regulation; (3) the red light turns on before the vehicle crosses, and it proceeds—violating the red-light regulation. Despite representing different legal outcomes, these scenarios share an identical spatial region of behavioral induction. Only the infrastructure state (i.e., signal status) differentiates the violation outcomes.
\begin{figure*}[htbp]
    \centering
    \includegraphics[width=0.85\textwidth]{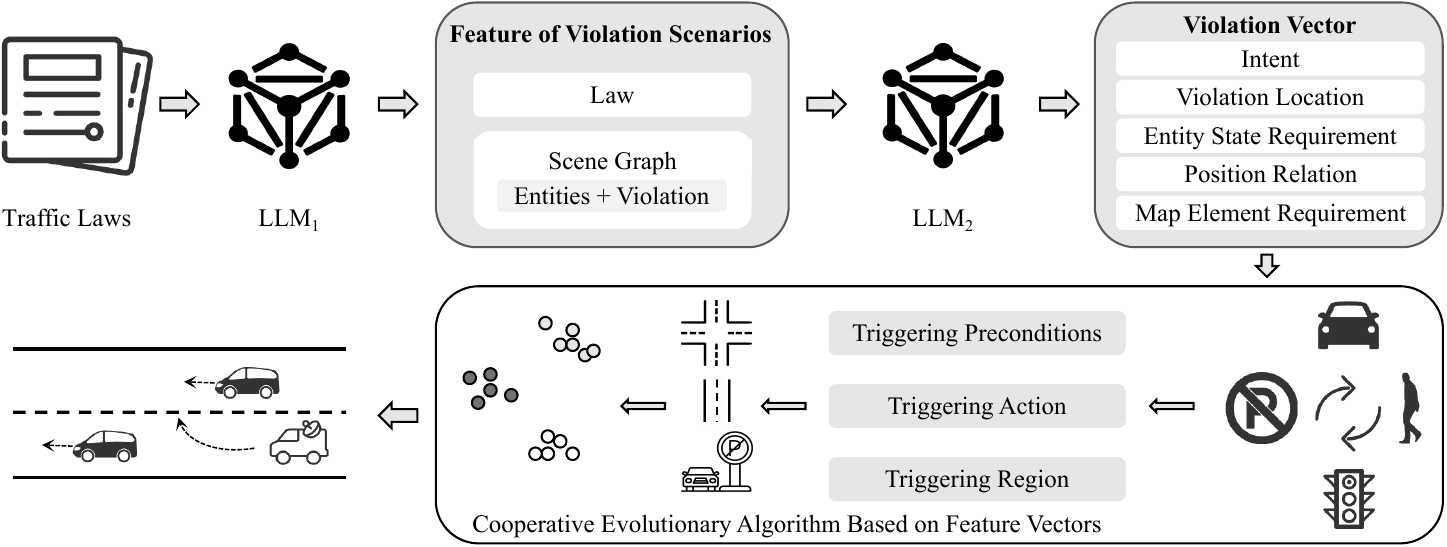}
    \caption{Overall framework of SLaFE}
    \label{fig:flow}
\end{figure*}

Motivated by this observation, we propose a unified framework for traffic law violation modeling and testing based on region semantics and language model reasoning. Leveraging LLMs' semantic understanding capabilities to extract key elements from regulations and cluster semantically similar rules, our approach employs co-evolutionary fuzzing across these clusters, enabling more efficient, generalizable, and scalable ADS compliance testing.

\section{Methodology}

In this section, we present the design details of SLaFE, a framework that leverages LLMs to vectorize traffic regulations and enhance the efficiency of ADS testing through a co-evolutionary algorithm based on feature vectors. The workflow of SLaFE is presented in Fig.~\ref{fig:flow}. 

The SLaFE workflow begins by using a large language model LLM$_1$ to extract semantic features from traffic regulation texts, identifying potential violation conditions (Section~\ref{sub:b}). These features are then encoded into structured violation feature vectors via a second model LLM\textsubscript{2} (Section~\ref{sub:c}). Based on these vectors, SLaFE generates and categorizes candidate initial scenes, and applies a co-evolutionary algorithm to evolve scenarios that effectively induce autonomous vehicles (AVs) to violate the corresponding laws (Section~\ref{sub:d}).

\subsection{Violation Modeling Based on Regional Semantics}\label{sub:a}

To enable interpretable traffic violation detection in ADS, we present a region-semantic modeling framework that transforms complex traffic laws into a structured and machine-readable form. Specifically, we decompose the semantics of traffic laws into two key components: (1) triggering preconditions, which capture time-sensitive conditions under which a rule becomes applicable, and (2) spatial violation constraints, which define location-specific behavioral restrictions. This decomposition enables unified modeling and interpretation of violation scenarios. Our approach introduces the following key improvements in terms of modeling granularity and semantic expressiveness: 

\begin{itemize}
\item \textbf{Intention-Free Rule Modeling.} Unlike VioHawk, which relies on intention-condition pairs to identify violation regions, our approach defines violations directly through observable behaviors and spatial relations. This removes redundant intention modeling, simplifies rule representation, and treats intention as an auxiliary feature for post-hoc interpretation or analysis.
\item \textbf{Primitive-Based Rule Modeling.} To better represent heterogeneous traffic laws, we design a unified set of well-defined and semantically consistent primitives that standardize map elements, traffic entities, and their states. This resolves inconsistencies and granularity mismatches in rule definitions, enabling more systematic and composable modeling.
\end{itemize}

\begin{table}[htbp]
  \centering
  \scriptsize
  \caption{Constituent elements of a test scenario}
  \label{tab:Element}
  \begin{tabularx}{\columnwidth}{>{\raggedright\arraybackslash}p{0.16\linewidth} 
                                   >{\raggedright\arraybackslash}p{0.16\linewidth} 
                                   X}
    \toprule
    \textbf{Category} & \textbf{Symbol} & \textbf{Definition} \\
    \midrule
    Map Structure        & I, L, C, T, SL, DYL & Intersection, Lane, Crosswalk, Traffic Light, Stop Line, Double Yellow Line \\
    Traffic Entity       & EGO, NPC, PED       & Ego Vehicle, Non-Ego Vehicle, Pedestrian \\
    Spatial Relationship & N, A                & Near, At \\
    Dynamic Attribute    & H, V                & Heading, Velocity \\
    \bottomrule
  \end{tabularx}
\end{table}

\noindent\textbf{Symbol Definition.}
We introduce a set of structurally consistent symbols to describe the key elements in a driving scenario, including the map structure, traffic entities, and their corresponding state conditions, as summarized in Table~\ref{tab:Element}.

\noindent\textbf{Triggering Preconditions.}
A triggering precondition refers to the combination of map states and entity states that satisfy the activation conditions of a traffic law at a specific time \textit{t}. We formalize these conditions as time-sensitive state expressions to support the temporal modeling and dynamic detection. For example, consider the traffic law: ``Vehicles must not enter a crosswalk while a pedestrian is crossing.''. The corresponding triggering preconditions include the following two elements:

\begin{itemize}
  \item \textit{Map Structure:} A crosswalk region labeled \( C_i \) exists in the current map, denoted as
  \[\
    Area(C_i).exist = True
  \]
  
  \item \textit{Entity State:} At least one pedestrian is present within region \( C_i \) and is currently walking, expressed as:
  \[
    PED \in Area(C_i) \wedge PED.state = WALK
  \]
\end{itemize}

\noindent\textbf{Behavioral Constraints.} A behavioral constraint defines the condition under which a specific action, when performed under satisfied triggering preconditions, results in the violation of a traffic law. It is jointly determined by a triggering action and a triggering region. Taking the pedestrian-priority rule as an example: when a pedestrian is actively crossing, the ego vehicle into the crosswalk constitutes a violation. The behavioral constraint in this case can be expressed as: $V[EGO,C_i]@[t_0,t_0+\Delta t]>\epsilon\wedge A(EGO,C_i)@[t_1]$. Where $V[EGO,C_i]@[t_0,t_0+\Delta t]$ denotes the component of the ego vehicle’s velocity directed toward crosswalk $C_i$ over the time interval $[t_0,t_0+\Delta t]$, indicating its intention to proceed; $A(EGO,C_i)@[t_1]$ indicates that the ego vehicle has entered region $C_i$ at time $t_{1}$.

\noindent\textbf{Unified Rule Modeling.} Based on the above-defined primitives and expression mechanisms, we propose a unified modeling format for traffic rules:\[
\resizebox{\linewidth}{!}{
  $\textbf{IF} \;
  \left[ \text{Precondition}_1 \land \text{Precondition}_2 \land \cdots \right]
  \;\Rightarrow\;
  \left[ \text{ViolationCondition} \right]$
}
\]

Using the pedestrian-priority rule as an example, the traffic rule can be formalized by combining its triggering preconditions and behavioral constraints as:
{\small
\[
\begin{aligned}
& \underbrace{Area(C_i).exist}_{\text{Map Structure}} \wedge \\
& \underbrace{(PED.state = WALK \wedge EGO \in Area(C_i) \wedge V(EGO) > \epsilon)}_{\text{Entity State}} \\
& \quad \Rightarrow Violation(\text{YieldToPedestrian})
\end{aligned}
\]
}

This unified representation balances generality and extensibility. Its primitive-based structure ensures semantic clarity and formal consistency, while maintaining adaptability to various traffic law modeling tasks. The consistent format supports scalable and interpretable rule encoding across diverse autonomous driving scenarios.

\subsection{Feature Extraction of Violation Scenarios}\label{sub:b}

To support effective regulation vectorization, we design a two-stage LLM framework. The first stage focuses on extracting structured semantic features from natural-language traffic laws, capturing the conditions under which violations may occur. This helps mitigate ambiguity common in free-text modeling by enforcing a structured intermediate representation, thereby improving accuracy and consistency in downstream processing.

To ensure the diversity of traffic regulations, we select representative regulatory texts from both China and the United States as modeling targets. These regulations specify the behavioral constraints that vehicles must follow under specific traffic structures and contextual conditions. The selected rules span a variety of road configurations, including intersections (with traffic signals and pedestrian crosswalks), multi-lane roads, and stop-sign zones.

The first-stage LLM (denoted as LLM\textsubscript{1}) produces a structured output called the Characteristics of Violation Scenarios (CVS). Each CVS is a tuple \textless{}\textit{Law}, \textit{Scene\_Graph}\textgreater{}, linking a specific regulation with its potential violation context.

\begin{itemize}
  \item \textit{Law} denotes the legal statement in natural language.
  \item \textit{Scene\_Graph} (SG) is a structured representation that captures the semantic conditions of violation scenarios, including traffic entities, spatial relations, and the associated violations.
\end{itemize}

\begin{figure}[htbp]
\centering
\begin{minipage}{0.85\linewidth}
\begin{lstlisting}[
basicstyle=\ttfamily\tiny,
keywordstyle=\bfseries,
morekeywords={<Scene_Graph>,<Entities>,<Violation>} 
]
<Scene_Graph> ::= <Entities>; <Violation>
<Entities> ::= <Id>; <Type>; <State>
<Id> ::= EgoVehicle|StopSign|Intersection|...
<Type> ::= VEHICLE|PEDESTRIAN|TRAFFIC_LIGHT | ...
<State> ::= <Motion>; <Spatial_Relationship>; <Attributes>
<Motion> ::= MOVING | STOPPED | NULL
<Spatial_Relationship> ::= Is_at | Is_next | Is_cross | ...
<Attributes> ::= RED | GREEN | YELLOW
<Violation> ::= <Description>; <Violating_Action>; <Violating_Entity>
<Description> ::= [Description of violation of regulations]
<Violating_Action> ::= [Actions that violate regulations]
<Violating_Entity> ::= EgoVehicle
\end{lstlisting}
\end{minipage}
\caption{The syntax of \textit{Scene\_Graph} DSL.}
\label{fig:scenegraph-dsl}
\end{figure}

To ensure consistency and parsability, the SG is specified using a DSL format. The structure of this representation is presented in Fig.~\ref{fig:scenegraph-dsl}.

This DSL is primarily inspired by the scenario generation approach proposed in \cite{sun2022lawbreaker}, and is designed to explicitly characterize violation-centric traffic scenarios. The core structure of the SG consists of two main components: \textit{Entities} and \textit{Violation}, each of which is further composed of sub-components and atomic elements, as detailed below.

\begin{itemize}
  \item \textit{Entities} represent all traffic participants involved in the violation scenario. Each entity is defined by a unique identifier (\textit{Id}), an entity type (\textit{Type}), and a set of state attributes (\textit{State}). The State component captures three aspects: the motion status of the entity (\textit{Motion}), its spatial relationship with specific road structures (\textit{Spatial\_Relationship}), and violation-relevant properties (\textit{Attributes}).
  \item \textit{Violation} specifies the conditions under which traffic regulations are breached. It includes a textual description of the violated regulation (\textit{Description}), a natural language description of the violating behavior (\textit{Violating\_Action}), and the traffic entity responsible for the violation (\textit{Violating\_Entity}). Both \textit{Description} and \textit{Violating\_Action} are expressed in natural language, as they are primarily intended to serve as inputs for feature extraction by the downstream language model component.
\end{itemize}

For each sub-component, we define a set of atomic elements to represent commonly observed attributes in violation scenarios. For example, under \textit{Spatial\_Relationship}, elements such as \textit{Is\_at}, \textit{Is\_next}, or \textit{Is\_cross} capture spatial predicates. To improve entity state accuracy, the values in the \textit{Entities} component are drawn from predefined ranges, whereas the fields in the \textit{Violation} component are generated freely by the LLM$_1$ based on its semantic interpretation of the regulation.

\noindent\textbf{Prompt Engineering.}
The generation of CVS relies heavily on well-designed prompt engineering. To enable the large language model to semantically interpret regulatory texts in a structured manner, we design a tailored prompting strategy consisting of the following components:

\begin{itemize}
  \item \textbf{Role Assignment.} The model is prompted to act as a \textit{Driving Violation Analyst}.
  \item \textbf{Task Definition.} The task is clearly stated as extracting and structuring violation scenarios from regulatory texts.
  \item \textbf{Output Schema Specification.} The expected output format is explicitly defined, detailing the semantics and type constraints of each field in the CVS structure.
  \item \textbf{Exemplar Demonstration.} Example inputs and outputs are provided to illustrate the expected structure and guide the model’s understanding.
\end{itemize}

The following instruction is provided to LLM\textsubscript{1} for extracting the CVS:
\begin{tcolorbox}[
    colback=gray!5,
    colframe=gray!60,
    title=Prompt for \textit{LLM}$_1$,
    fonttitle=\bfseries,
    coltitle=black,
    boxrule=0.4pt,
    arc=2mm,
    left=1mm,
    right=1mm,
    top=0.5mm,
    bottom=0.5mm
]

\textbf{Role:} You are a \textit{Driving Violation Analyst}.\\[0.5ex]
\textbf{Task:} Your task is to analyze a traffic law and deconstruct the exact moment a violation occurs into a structured JSON object.\\[0.5ex]
\textbf{CVS Format:}
\begin{itemize}
\item \textbf{Entities:} A list of entities involved at the moment of the violation.

\item \textbf{Violation:} An object describing the violation event.

\end{itemize}
\textbf{Example:} \textit{...}

\end{tcolorbox}

Crucially, the intermediate representation generated in this stage serves as the direct contextual input for the subsequent violation vector construction. This design ensures the semantic consistency and controllability of traffic laws across the two stages.

\subsection{Violation Feature Vector Conversion}\label{sub:c}

The second-stage LLM(denoted as LLM\textsubscript{2}) converts the structured CVS output into a Violation Feature Vector. The VFV conversion stage serves as a bridge between the CVS and the fuzz testing engine. It refines the structured features of a violation scenario, transforming them into a set of normalized, traffic-rule-aligned conditional variables. These variables are specifically engineered to steer the co-evolutionary fuzzing process.

To structurally represent the conditions for triggering a traffic rule violation, we define the violation vector as a 5-tuple:
\textless{}\textit{Intent}, \textit{Violation\_Location}, \textit{Entity\_State\_Requirements}, \textit{Position\_Relation}, \textit{Map\_Element\_Requirement}\textgreater{}.

\begin{itemize}
  \item \textit{Intent} describes the ego vehicle’s high-level driving intention at the time the violation occurs.

  \item \textit{Violation\_Location} denotes the specific map region where the violation takes place.

  \item \textit{Entity\_State\_Requirement} specifies the state conditions of surrounding traffic participants that must be met for the violation to be triggered.

  \item \textit{Position\_Relation} captures the spatial relationship between the vehicle and a critical reference point related to the violation.

  \item \textit{Map\_Element\_Requirement} defines the existence and attributes of relevant map structures (e.g., crosswalks, stop lines).
\end{itemize}

\noindent\textbf{Prompt Engineering.}
To address potential issues in multi-task extraction using LLMs, such as inaccurate field identification or violations of value constraints, we decompose the vector construction process into five interrelated but modular subtasks. For each variable, we design dedicated prompts that leverage both the contextual information and previously extracted values. For example, when extracting \textit{Entity\_State\_Requirement}, the prompt explicitly instructs the model to focus on the \textit{Entities} field in the violation scenario representation to ensure consistency and accuracy.

The following is a sample prompt template used for extracting entity state requirements:

\begin{tcolorbox}[
    colback=gray!5,
    colframe=gray!60,
    title=Prompt for \textit{LLM}$_2$,
    fonttitle=\bfseries,
    coltitle=black,
    boxrule=0.4pt,
    arc=2mm,
    left=1mm,
    right=1mm,
    top=0.5mm,
    bottom=0.5mm
]

\textbf{Role:} You are analyzing a structured traffic violation scene.\\[0.5ex]
\textbf{Task:} Given the CVS below, extract the following attribute.\\[0.5ex]
\textbf{Attribute:} Entity\_State\_Requirement\\[0.5ex]
\textbf{Value Range \& Description:} Define the expected motion state, spatial relationship, or relevant attributes for each entity involved in the violation.\\[0.5ex]
\textbf{ATTENTIONS:} A1, A2, A3, \text{...}.
\end{tcolorbox}

It is important to note that the ``Value Range \& Description'' section specifies the allowable values for each variable, along with a concise description of their semantics. Additionally, the ``ATTENTIONS'' section provides guidance for LLM\textsubscript{2} during the conversion process, highlighting contextual dependencies between the current target variable and previously extracted variables.

\subsection{Cooperative Evolutionary Algorithm Based on Feature Vectors}\label{sub:d}

Upon completing the two-stage LLM-based feature extraction process, we obtain structured VFVs representing traffic violations. Building on these vectors, we propose a cooperative evolutionary algorithm guided by feature vectors, which aims to alleviate the cost of manually constructing initial scenarios and improve the efficiency of discovering violation-inducing cases in high-dimensional spaces.

\noindent\textbf{Violation Trigger Modeling.}
To bridge natural language traffic regulations with scenario generation, we propose a violation trigger modeling mechanism that converts structured VFVs into region-level constraints for scenario construction. Specifically, we define a traffic violation in autonomous driving as the act of entering a legally restricted area under specific map structures and dynamic context conditions. This formalization provides a clear optimization objective for the evolutionary algorithm, while enabling the extraction of concrete, spatially-grounded violation triggers from abstract regulatory semantics.

Specifically, we construct the violation-triggering conditions from the following three aspects:

\begin{itemize}
    \item \textbf{Triggering Preconditions:}  
    When the \textit{entity\_state\_requi-\\rements} refer to non-ego participants, they are combined with \textit{position\_relation} and \textit{violation\_location} to capture the spatial and contextual constraints imposed by these entities. Together with the \textit{map\_element\_requirement}, these components jointly define the triggering preconditions of a violation.

    \item \textbf{Triggering Action:}  
    The intent specifies the planned maneuver of the ego vehicle. When combined with an \textit{entity\_state\_requirement} that constrains the ego's own state, it jointly defines the triggering action that may lead to a violation.

    \item \textbf{Triggering Region:}  
    Combining \textit{position\_relation} and \textit{violation\_location} specifies the spatial region associated with the violation (e.g., “on a crosswalk”, “near a stop sign”), which we refer to as the triggering region.
\end{itemize}

If the starting field of any combination is labeled NONE, the regulation does not impose explicit requirements on that aspect, and the combination is excluded from violation modeling.

Through this structured modeling pipeline, each regulation is systematically translated into constraint-based triggers that form the basis for region-guided scenario generation and spatial mutation.

\noindent\textbf{Scene Configuration Generation.}
Given the diversity of traffic regulations and the variations across different countries, we propose an initial configuration generation method following the violation trigger modeling phase. This method leverages feature vectors as conditional constraints to systematically construct effective initial scenarios without preset traffic elements.

Our approach adopts a stepwise configuration strategy, initializing critical map elements and entity states based on the violation conditions specified in the feature vector. The vehicle’s start and end positions are first determined by the \textit{violation\_location} field, ensuring that its nominal trajectory traverses areas prone to violations.

Subsequently, essential map structures are instantiated according to the \textit{intent} and \textit{map\_element\_requirement} fields. For instance, traffic signals follow a standard red–yellow–green cycle when the intent is \textit{PASS\_INTERSECTION}, remain green throughout for \textit{PARK}, and are omitted entirely for \textit{CHANGE\_LANE}. If the \textit{map\_element\_requirement} field explicitly specifies traffic signal states, these override the default configurations.

Finally, the initial states of all relevant traffic participants (e.g., pedestrians in motion, stationary vehicles) are established in accordance with the \textit{entity\_state\_requirement} attribute, thereby satisfying the preconditions required to induce the target violation scenarios.

The scenario configurations generated at this stage serve as the initial seeds for collaborative evolution. 

\begin{algorithm}[htbp]
\begin{spacing}{1.2}
\small
\caption{Regional Guided Mutation}
\label{alg:regional-guided-mutation}
\KwIn{Scenario $s$}
\KwIn{Entity List $e$}
\KwIn{Frame Time $t$}
\KwOut{Mutated Scenario $s'$}

\SetKwFunction{FMutate}{Regional\_Guided\_Mutate}
\SetKwProg{Fn}{Function}{:}{}
\Fn{\FMutate{$s$, $t$, $e$}}{

    $s' \gets s$\;
    
    $R_{\text{max}} \gets 0$\;

    $P_{\text{max}} \gets 0$\;
    
    $ego \gets$ \texttt{Get\_Ego\_Position}$(s, t)$\;

    $grid \gets$ \texttt{Creat\_Grid}$(ego)$\;

    \ForEach{$partition$ $p \in grid$}{
        $R_D \gets$ \texttt{Cal\_Drivable\_Region}$(s, t)$\;
        
        $R_T \gets$ \texttt{Get\_Trigger\_Region}$(s, t)$\;
        
        $R_{DNT} \gets R_D \setminus R_T$\;

        \If{$R_{\text{DNT}} > R_{\text{max}}$}{
            
            $R_{\text{max}} \gets R_{\text{DNT}}$\;
            
            $P_{\text{max}} \gets p$\;
        }
    }

    \If{\texttt{Is\_Path\_Blocked}$(e$, $ego$, $grid$) = $\text{False}$}{
        $s' \gets$ \texttt{Shrink\_DNT\_Area}($s'$, $P_{\text{max}}$)\;
    }
    \Else{
        $s' \gets$ \texttt{Expand\_TND\_Area}($s'$, $e$, $ego$, $grid$)\;
    }

    \KwRet{$s'$}
}
\end{spacing}
\end{algorithm}

\noindent\textbf{Regional Guided Mutation.}
Following the initial configuration generation, the evolutionary algorithm requires effective mutation mechanisms to continuously uncover potential violations. However, since violations are often contingent on specific geographic locations and traffic configurations, conventional random mutations tend to be inefficient. To address this challenge, we propose a regional guided mutation strategy (listed in Algorithm \ref{alg:regional-guided-mutation}) designed to enhance the search efficiency within critical violation-prone areas.

This strategy reconstructs realistic violation behaviors in a structured manner based on regional semantics: the AV enters a predefined triggering region under the specific map and environmental conditions that would trigger the violation, thereby providing clear guidance for mutation directions. Specifically, the mutation process aims to constrain non-triggering areas where the vehicle can safely and legally drive, while expanding the potential path space leading toward the triggering region.

To constrain non-triggering regions, we draw inspiration from the mutation heuristics proposed in VioHawk. Specifically, we construct a 3×3 grid centered on the AV and apply the drivable area estimation method from \cite{soentges2017computing} to calculate the vehicle's reachable space under its current motion state. Based on the spatial distribution of non-triggering regions within this area, we adopt a directional mutation strategy to restrict the vehicle's motion path and suppress exploration toward irrelevant regions.

Conversely, to improve access to triggering regions, we evaluate the spatial relationship between surrounding traffic participants and the violation zone. If these regions are obstructed and the vehicle is unable to reach them, we apply adaptive behavior adjustments to nearby entities. These adjustments expand the drivable area and facilitate the ego vehicle's access to the designated violation region.

\begin{figure}[htbp]
    \centering
    \includegraphics[width=0.48\textwidth]{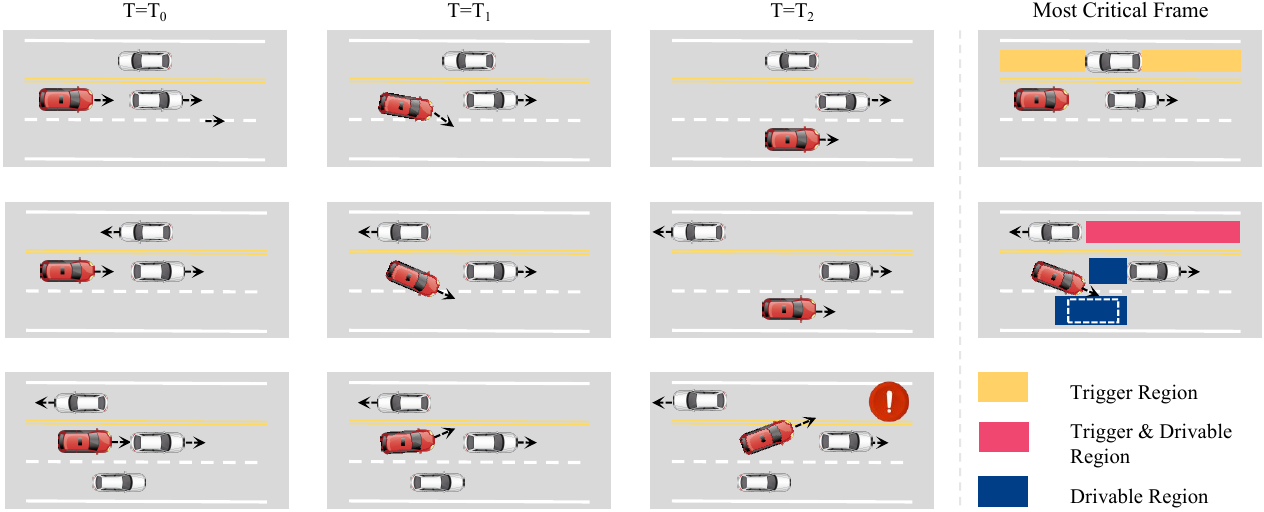}  
    \caption{Example of regional guided mutation}
    \label{fig:region}
\end{figure}

As shown in Fig.~\ref{fig:region}, regional guided mutation first expands the drivable triggering region by mutating the speed of vehicles located within it. Then, based on a grid-based partitioning of the scene, additional vehicles are introduced via mutation in the front-right non-triggering area. This placement constrains the non-triggering regions, thereby guiding the ego vehicle toward the violation zone and increasing the likelihood of regulation violations.

\begin{algorithm}[htbp]
\setcounter{AlgoLine}{0}
\begin{spacing}{1.2}
\small
\caption{Similarity Cooperative Evolution}
\label{alg:similarity_cooperative_evolution}
\KwIn{Regulation List $L$}
\KwIn{Scene Parameter Vectors $V$}
\KwOut{Violation-inducing Scenarios $S$}

\SetKwFunction{FEvolution}{Sim\_Cooperative\_Evolution}
\SetKwProg{Fn}{Function}{:}{}
\Fn{\FEvolution{$L$, $V$, $S$}}{

    $G \gets$ \texttt{Categorize\_Regulations }($L$, $V$)\;
    
    $P \gets$ \texttt{Initialize\_Population}($L$, $G$)
    
        \ForEach{$law \in L$}{
                $seed \gets \texttt{Get\_Seed\_From}(law)$\;
            
                $candidate \gets \texttt{Region\_Guided\_Mutate}(seed)$\;

                $trace \gets \texttt{Run\_Instance}(candidate)$\;

                \If{$candidate$ violates $law$}
                {
                    add $candidate$ to $S$
                }
                
                $scores, mutators \gets \texttt{Evaluate}(candidate, trace)$\;

                \eIf{$\texttt{Not\_Full}(law)$}{
                    \texttt{Add\_To\_Pop}$(P[law], candidate, mutators)$\;
                }{
                    \If{$\texttt{Is\_Better\_Seed}(law, scores)$}{
                        \texttt{Replace}$(P[law], candidate, mutators)$\;
                    }
                }
                $group = G[law]$
                
                \ForEach{$law' \in group \setminus \{law\}$}{
                    \If{$\texttt{Is\_Better\_Seed}(law', scores)$}{
                        \texttt{Replace}$(P[law], candidate, mutators)$\;
                    }
                }
        }
    \KwRet{S}
}
\end{spacing}
\end{algorithm}

\noindent\textbf{Similarity Cooperative Evolutionary.}
To enable concurrent evaluation of multiple traffic regulations and improve the efficiency of discovering violation-inducing scenarios, we propose a similarity-based cooperative evolutionary strategy (listed in Algorithm \ref{alg:similarity_cooperative_evolution}). This strategy is motivated by a key observation: although different regulations may guide violations in distinct triggering regions, evolving them independently leads to low overall efficiency. However, if we can effectively identify structural similarities between regulations, high-value scenarios can be shared across them, significantly enhancing the overall evolution process.

Given the inherent semantic ambiguity and syntactic diversity of traffic regulations expressed in natural language, directly measuring similarity at the textual level is often unreliable. To overcome this limitation, we introduce structured scene parameter vectors as abstract representations of regulatory constraints. These vectors are derived from the configuration options generated during the scene configuration generation phase, and capture the essential features required for triggering each regulation. By leveraging these vectors, we group regulations with similar structural characteristics into co-evolutionary subgroups, which serve as the fundamental units for our coordinated evolutionary process.

To quantify the potential of a scenario to trigger a violation, we define the critical score as:

\[
Critical\_score = \frac{|R_D \cap R_T|}{|R_D|}
\]

Here, \(R_D\) denotes the drivable region of the AV in the current frame, and \(R_T\) represents the violation-triggering region. 
The intersection \(R_D \cap R_T\) indicates the drivable portion that overlaps with the triggering region.

This score quantifies the ``violation potential'' of the scenario under a given regulation: the higher the score, the closer the vehicle is to entering the violation-triggering zone.

During each evolutionary round, we maintain a dedicated main population for each regulation, responsible for producing locally optimized violation scenarios. All candidate seeds generated by a main population are retained unconditionally. Once the population exceeds a predefined size threshold, we perform selection based on their criticality scores, ensuring that high-potential scenarios are prioritized.

Meanwhile, each candidate seed is also evaluated for its criticality under other regulations within the same co-evolutionary subgroup—i.e., those categorized by structural similarity in scene parameter vectors. If a seed demonstrates high violation potential for another regulation, it is transferred together with its associated mutator into the main population of that law.

This cross-subgroup sharing mechanism facilitates cooperative evolution among structurally similar regulations, promoting the reuse of high-quality seeds while preserving the stability and optimization efficiency of individual populations.

\begin{table*}[htbp]
  \centering
  \small
  \caption{Selected traffic rules}
  \label{tab:rules}
  \begin{tabularx}{\textwidth}{
    >{\centering\arraybackslash}p{0.15cm}
    >{\centering\arraybackslash}p{1.68cm}
    >{\raggedright\arraybackslash}c
  }
    \toprule
    \textbf{ID} & \textbf{Data Source} & \textbf{Description} \\
    \midrule
    R$_1$ & Chinese Law & When the red light is on, vehicles can not cross the intersection. \\
    R$_2$ & Chinese Law & When the yellow light is on, only vehicles that have crossed the stop line can cross the intersection. \\
    R$_3$ & Chinese Law & When overtaking, vehicles shall not affect driving of the vehicles in the adjacent lane. \\
    R$_4$ & U.S. Law & When a vehicle has stopped behind the crosswalk, vehicles shall not cross the crosswalk. \\
    R$_5$ & U.S. Law & When a pedestrian is at the crosswalk, vehicles shall not pass the crosswalk. \\
    R$_6$ & U.S. Law & Vehicles shall not park at the crosswalk. \\
    R$_7$ & U.S. Law & Vehicles shall not park within 20 feet of a crosswalk. \\
    R$_8$ & U.S. Law & Vehicles shall not park within 30 feet of a traffic light. \\
    R$_9$ & U.S. Law & Vehicles shall not park within 30 feet of a stop sign. \\
    R$_{10}$ & U.S. Law & When a pedestrian is on the crosswalk at the turning side of the intersection, vehicles shall not enter the intersection. \\
    \bottomrule
  \end{tabularx}
\end{table*}

\begin{table*}[htbp]
\centering
\caption{The time required to trigger the traffic violation}
\label{tab:time_trigger}
\scriptsize
\begin{tabularx}{\textwidth}{c *{7}{>{\centering\arraybackslash}X}}
\toprule
\textbf{ID} & \textbf{AV-Fuzzer} & \textbf{DriveFuzz} & \textbf{LawBreaker} 
& \textbf{ABLE} & \textbf{AutoFuzz} 
& \textbf{VioHawk} & \textbf{SLaFE} \\
\midrule
R-1  & \ding{55} & \ding{55} & 33m53s & 13m32s & 12m58s & 10m10s & 3m24s \\
R-2  & \ding{55} & \ding{55} & 24m12s & 26m24s & \ding{55} & 3m19s & 2m35s \\
R-3  & \ding{55} & \ding{55} & \ding{55} & \ding{55} & \ding{55} & 20m57s & 20m28s \\
R-4  & \ding{55} & \ding{55} & \ding{55} & 27m41s & \ding{55} & \ding{55} & 2m11s \\
R-5  & \ding{55} & \ding{55} & \ding{55} & \ding{55} & \ding{55} & 5m22s & 3m29s \\
R-6  & \ding{55} & \ding{55} & \ding{55} & \ding{55} & \ding{55} & 9m25s & 1m46s \\
R-7  & \ding{55} & \ding{55} & \ding{55} & \ding{55} & \ding{55} & 6m47s & 7m39s \\
R-8  & \ding{55} & \ding{55} & \ding{55} & \ding{55} & \ding{55} & 14m3s & 1m20s \\
R-9  & \ding{55} & \ding{55} & \ding{55} & \ding{55} & \ding{55} & 3m29s & 3m26s \\
R-10 & \ding{55} & \ding{55} & \ding{55} & \ding{55} & \ding{55} & 7m22s & 4m21s \\
\bottomrule
\end{tabularx}
\end{table*}

\section{Evaluation}
We conducted a series of experiments to evaluate the performance of the proposed method, addressing the following three research questions:

RQ1: How effective is SLaFE in identifying scenarios that violate traffic laws?

RQ2: How efficient is the cooperative evolutionary algorithm in SLaFE at discovering law-violating scenarios?

RQ3: How accurate are LLM\textsubscript{1} and LLM\textsubscript{2} for the transformation tasks in SLaFE?

\subsection{Experimental Setup}

\noindent\textbf{Target ADS.}
We adopt Baidu Apollo as the target ADS, as it is an open-source, industrial-grade platform that supports a wide range of driving functionalities. Apollo has been widely used in prior ADS-related research as a benchmark system\cite{sun2022lawbreaker, tang2024legend, li2024viohawk}. In our experiments, we focus on the stable version, Apollo 7.0.

\noindent\textbf{Target Simulation Platform.}  
We utilize LGSVL as the simulation platform due to its high-fidelity simulation capabilities and seamless integration with Apollo. LGSVL is open-source and offers strong extensibility, making it a popular choice for evaluating ADS behavior. We use version LGSVL 2021.2.2 in our study.

\noindent\textbf{Baselines.}  
We compare SLaFE with six representative baselines drawn from prior work. Among them, LawBreaker~\cite{sun2022lawbreaker}, ABLE~\cite{zhang2023testing}, and VioHawk~\cite{li2024viohawk} are state-of-the-art tools designed to detect complex traffic law violations in real-world scenarios. In addition, we include AV-Fuzzer~\cite{li2020av}, DriveFuzz~\cite{kim2022drivefuzz}, and AutoFuzz~\cite{zhong2022neural}, representative fuzzing frameworks focusing on either collision-triggering violations or basic rule violations.

\noindent\textbf{Configurations.}  
SLaFE employs GPT-4o-mini for both the \emph{feature extraction of violation scenarios} and the \emph{violation feature vector conversion} stages, as this model offers a good trade-off between accuracy and inference latency. We set the temperature to 0.2 to balance response determinism with the reasoning flexibility required for violation interpretation.

All experiments were conducted on a desktop machine running Ubuntu 20.04, equipped with an Intel(R) Xeon(R) Gold 6226R CPU, 64GB RAM, and an Nvidia RTX 3090 GPU. This hardware configuration ensures sufficient computational resources and stability for running both the ADS and simulation platform. For testing scenarios, we selected the San Francisco HD map provided by LGSVL, as it offers rich road structures suitable for evaluating diverse traffic laws.

\subsection{\textbf{RQ1: How effective is SLaFE in identifying scenarios that violate traffic laws?}}

\noindent\textbf{Experiment Design.} To comprehensively evaluate the performance of SLaFE in discovering law-violating scenarios, we conducted a comparative study focusing on two key aspects: violation type coverage and violation triggering efficiency. The experiment included ten representative traffic laws selected from China and the United States (see Table \ref{tab:rules}). 

Since all baseline methods rely on initial simulation seeds to explore violation scenarios, we provided each method with ten seeds aligned with the corresponding law-based tasks. Specifically, SLaFE automatically generated four types of seeds during the scenario configuration phase, which were then assigned to tasks based on law categories to ensure fairness and consistency. To control variables, we allocated a fixed exploration time of three CPU hours for each method.

\noindent\textbf{Experiment Results.}
As shown in Table~\ref{tab:time_trigger}, under the same time budget, SLaFE successfully triggered all 10 types of traffic law violations (10/10), demonstrating a clear advantage in violation discovery capability over existing methods. Among the baselines, the best-performing method, VioHawk, identified 9 types, while other approaches detected no more than 3. SLaFE also demonstrated a clear advantage in violation-triggering efficiency, discovering violations significantly faster than baseline methods across most law-based tasks, with an average improvement of approximately 54\%. These results further validate the effectiveness of SLaFE in identifying law-violating scenarios.

It is worth noting that SLaFE adopts a multi-task co-evolution strategy during exploration. To fairly evaluate the violation-triggering time for each law, we apply an inter-task differential calculation. Specifically, the triggering time for the i-th law is defined as the time difference between its triggering timestamp and that of the previously triggered law. This design allows us to assess the contribution of co-evolution in reducing marginal exploration cost, and enables a direct comparison with the triggering times of corresponding violations in baseline methods.

Table~\ref{tab:time_trigger} presents the violation triggering time required by each method for all individual law tasks. If a method fails to discover a violation scenario within the time limit, it is marked with “\ding{55}”. SLaFE outperforms VioHawk and other baselines in most tasks and exhibits particularly strong performance on R-4, R-6, and R-8—regulations that benefit from seeds evolved in related tasks, enabling faster violation discovery via cross-task behavioral transfer.

\begin{figure}[htbp]
    \centering
    \includegraphics[width=0.48\textwidth]{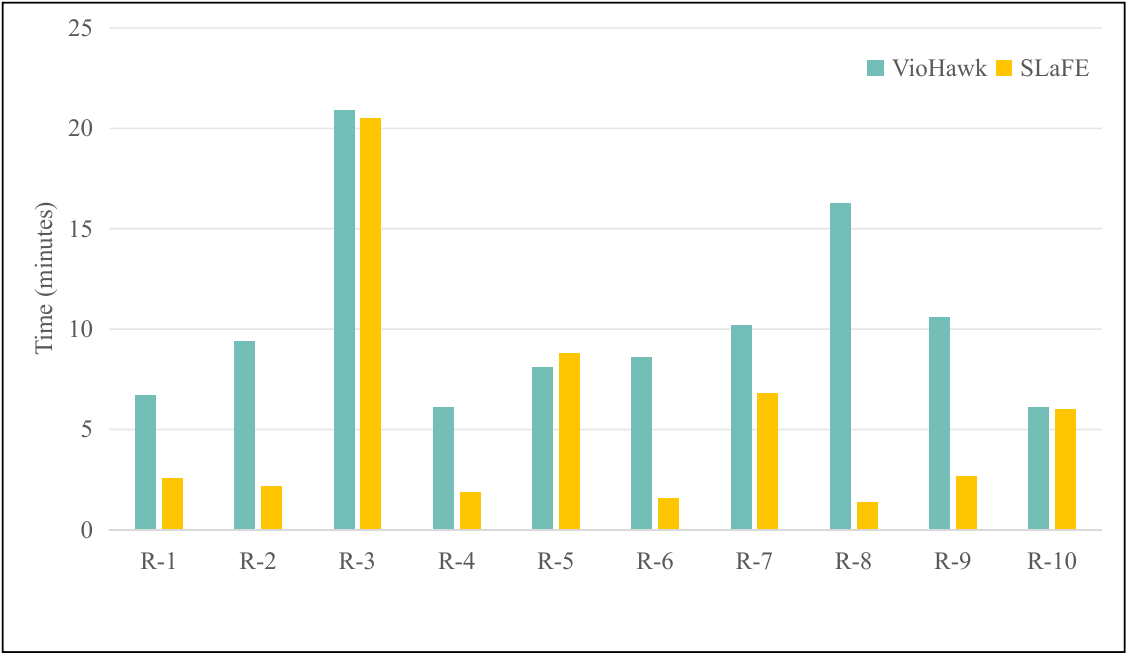}  
    \caption{Violation discovery efficiency of SLaFE and VioHawk }
    \label{fig:bar}
\end{figure}

To further evaluate the generalization capability of SLaFE, we conducted a comparative study against VioHawk on the traffic violation dataset\cite{li2024viohawk} (see Fig.~\ref{fig:bar}). In this setting, the initial scenarios were manually crafted for different traffic laws. Despite this, SLaFE was able to effectively transfer behavioral elements through its co-evolution strategy, resulting in a substantial improvement in violation-triggering efficiency.

\noindent\textbf{Analysis of Results.}
Beyond the quantitative results, we further analyze the co-evolutionary subgroups to explain SLaFE’s advantages. 

On regulations R-5 to R-10, which involve complex spatial semantics in parking and pedestrian interactions, the performance disparity across baselines is evident, exposing their respective limitations. VioHawk performs best by leveraging manually designed initial scenes and region-guided methods, but its efficiency still lags behind SLaFE. Lawbreaker and ABLE rank next, relying on formal languages that struggle to capture the full semantic complexity of these scenarios (with ABLE further aided by active learning). Other fuzzing methods almost completely fail, as they lack both formal traffic-law representations and deterministic mutation targets.

SLaFE’s improvements stem from the synergy of its components. The two-stage LLM process converts nuanced legal texts into precise VFV constraints, which, combined with region-guided mutation, yield targeted search directions. The cooperative evolution strategy further improves efficiency by sharing seeds across semantically related rules—for example, between pedestrian-at-crosswalk rules (R-5, R-10) and illegal parking rules (R-6–R-9). This mechanism also enhances robustness: on R-4, where VioHawk failed, SLaFE corrected deviations by leveraging seeds from similar rules. These factors, validated by our ablation study (Section~\ref{sub:e}), explain why SLaFE consistently outperforms baselines in challenging scenarios.

\subsection{\textbf{RQ2: How efficient is the cooperative evolutionary algorithm in SLaFE at discovering law-violating scenarios?}}\label{sub:e}

\noindent\textbf{Experiment Design.}
To quantitatively assess the impact of the cooperative evolutionary algorithm on violation discovery efficiency in SLaFE, we designed an ablation study. Specifically, we implemented a variant of SLaFE (denoted as SLaFE\_), in which each evolution round targets a single traffic law independently. This variant lacks seed sharing and behavioral transfer across law tasks. Both SLaFE and SLaFE\_ were provided with 10 initial simulation configurations matched to the corresponding traffic laws, and each task was allocated a uniform exploration time budget of 3 CPU hours.

\begin{table}[ht]
  \centering
  \scriptsize
  \caption{A comparison between SLaFE and SLaFE\_}
  \label{tab:comparison}
  
  \begin{tabularx}{\columnwidth}{l >{\centering\arraybackslash}X >{\centering\arraybackslash}X >{\centering\arraybackslash}X}
    \hline
    \textbf{Metric} & \textbf{TypeExpNum} & \textbf{SimForOneType} & \textbf{TimeForOneType} \\ \hline
    \textbf{SLaFE}      & 10/10               & 5.5                    & 5.1                     \\
    \textbf{SLaFE\_}     & 9/10                & 11.7                   & 8.6                     \\ \hline
  \end{tabularx}
\end{table}

\noindent\textbf{Experiment Results.}
As shown in Table~\ref{tab:comparison}, SLaFE\_ successfully triggered 9 out of 10 violation types within the allotted time. On average, it required 11.7 simulations per rule and 8.6 minutes to trigger a violation. In contrast, under the same settings, SLaFE successfully triggered all 10 violation types, with an average of 5.5 simulations and 5.1 minutes per rule. Compared to SLaFE\_, SLaFE achieves 53\% reduction in the number of simulations and 41\% reduction in triggering time.

\begin{table}[ht]
    \centering
    \begin{threeparttable}
        \caption{Time comparison between \textbf{SLaFE} and \textbf{SLaFE\_}}
        \label{tab:Time_comparison}
        \scriptsize
        \renewcommand{\arraystretch}{0.8}
        \setlength{\tabcolsep}{0.8pt}
        \begin{tabularx}{\linewidth}{
            >{\raggedright\arraybackslash}X
            @{\hspace{12pt}}
            *{11}{>{\centering\arraybackslash}X}}
            \toprule
            \textbf{ID} & 1 & 2 & 3 & 4 & 5 & 6 & 7 & 8 & 9 & 10 & \textbf{Total}\tnote{*} \\
            \midrule
            \textbf{SLaFE}   & 3.4 & 2.1 & 20.5 & 2.2 & 3.5 & 1.8 & 7.7 & 1.3 & 3.4 & 4.4 & 50.3 \\
            \textbf{SLaFE\_} & 3.4 & 9.8 & 20.9 & \ding{55} & 10.9 & 8.4 & 4.9 & 7.8 & 5.2 & 4.5 & 75.9 \\
            \bottomrule
        \end{tabularx}
        \begin{tablenotes}
            \tiny
            \item[*] The unit for all time values is minutes.
        \end{tablenotes}
    \end{threeparttable}
\end{table}

In addition, Table~\ref{tab:Time_comparison} compares the violation-triggering time of the two methods across all law tasks. For tasks where a method failed to trigger a violation, ``\ding{55}" is used as an indicator. As observed, SLaFE consistently outperforms SLaFE\_ in most tasks and exhibits faster convergence. These results highlight the effectiveness of the cooperative evolutionary algorithm in improving violation discovery efficiency and reducing search cost.

\noindent\textbf{Analysis of Failure Case.}
We further analyzed the failure case of SLaFE\_ on the R-4 law task. The law specifies that ``When a vehicle has stopped behind the crosswalk, vehicles shall not cross the crosswalk.''. In this task, SLaFE\_ performed independent exploration focused solely on the target law in each evolutionary round, without incorporating behavioral context from other tasks. As a result, the system failed to evolve the complex dynamic interactions required to violate the rule within the initial scenario, ultimately converging to a local optimum and failing to trigger the violation.

In contrast, SLaFE leverages its cooperative evolutionary algorithm to enable cross-task transfer of behavioral elements. During the evolution of the R-4 task, the system was able to utilize representative traffic scenarios previously evolved under traffic light-related laws. These scenarios provided the necessary behavioral preconditions to satisfy the triggering conditions of R-4, allowing the behavioral constraint to be effectively activated. This algorithm alleviates the exploration bottleneck faced by single-objective strategies when behavioral dependencies are not explicitly modeled, significantly improving both the efficiency and success rate of violation discovery.

\subsection{\textbf{RQ3: How accurate are $LLM_1$ and $LLM_2$ for the transformation tasks in SLaFE?}}

\noindent\textbf{Experiment Design.}
To evaluate the accuracy of LLM\textsubscript{1} and LLM\textsubscript{2} in performing law-to-scenario transformation within SLaFE, we designed two complementary evaluation methods: a user-based subjective survey and a feature-vector-based accuracy analysis.

For LLM\textsubscript{1}, we conducted an online user study involving ten participants from the field of computer science. The participants included four graduate students with over two years of research experience in autonomous driving and two professors with long-term expertise in system testing. Each participant was asked to read the traffic law texts and evaluate the CVS generated by LLM\textsubscript{1} on an item-by-item basis. The evaluation employed a 5-point Likert scale, ranging from “strongly disagree” to “strongly agree”. Participants assessed key fields including the \textit{entity} and \textit{violation}. They were also asked to briefly explain the rationale behind each rating to support subsequent qualitative analysis.

For LLM\textsubscript{2}, we adopted a dual-review protocol to evaluate the accuracy of feature vector extraction. Specifically, two researchers independently decomposed and aligned the feature vectors for each law. In cases of disagreement, they engaged in discussion to establish a consensus-based “human gold standard”, which was then used to calculate the accuracy of LLM\textsubscript{2}’s automatically extracted results.

\begin{figure}[htbp]
    \centering
    \includegraphics[width=0.48\textwidth]{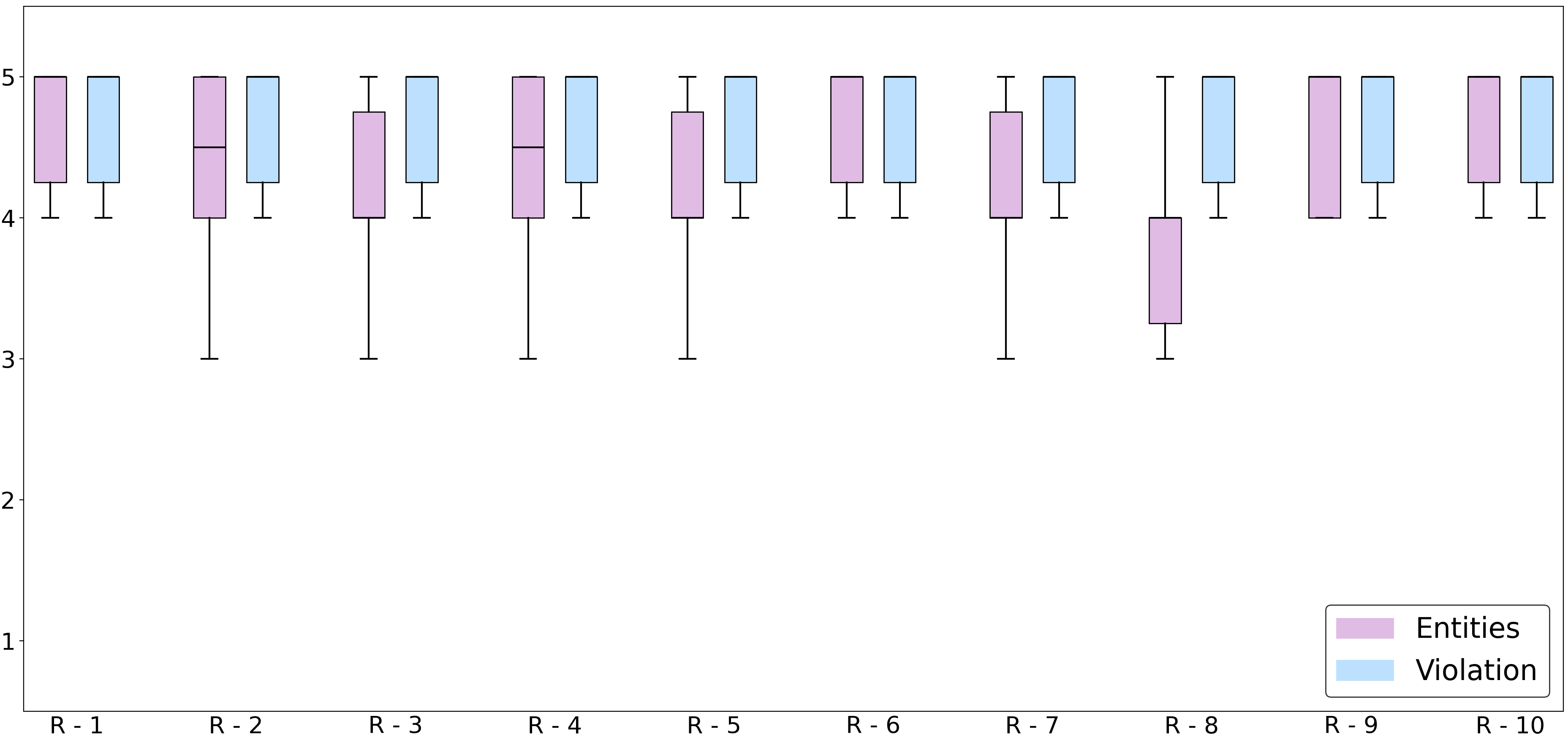}  
    \caption{Evaluation scores for 10 rules}
    \label{fig:ESC}
\end{figure}

\noindent\textbf{Experiment Results.}
Fig.~\ref{fig:ESC} presents user evaluations of the violation scenario features generated by LLM\textsubscript{1}. Overall, participants generally acknowledged the model’s effectiveness in translating traffic regulations into the structured components of the CVS. In particular, the transformation of regulatory descriptions into the \textit{violation} field was rated highly. A few participants noted ambiguity in the \textit{description} field, which typically stemmed from vagueness in the legal text itself. 

However, the \textit{entities} field—especially the representation of \textit{state} attributes—received relatively lower scores. Further analysis revealed that this was primarily due to the fact that some traffic laws do not explicitly specify the precise conditions or states of involved entities. In such cases, LLM\textsubscript{1} tends to infer and complete the missing information based on contextual cues, occasionally introducing irrelevant entities or incorrect states. 

Since the current transformation process relies heavily on the LLM’s backward reasoning capabilities, our ability to control the generated content remains limited. At this stage, we can only guide the output through prompt engineering, without enforcing strict constraints over the generation behavior.

\begin{table}[htbp]
\centering
\caption{Inquiry-stage configuration for each law}
\label{tab:accuracy}
\scriptsize
\renewcommand{\arraystretch}{1.2}
\begin{tabularx}{\linewidth}{c *{4}{>{\centering\arraybackslash}X}}
\toprule
\textbf{ID} 
& \textbf{\shortstack{One-stage\\One-inquiry}} 
& \textbf{\shortstack{One-stage\\Two-inquiry}} 
& \textbf{\shortstack{Two-stage\\One-inquiry}} 
& \textbf{\shortstack{Two-stage\\Multi-inquiry}} \\
\midrule
R-1  & \cmark & \xmark & \xmark & \cmark \\
R-2  & \xmark & \xmark & \xmark & \cmark \\
R-3  & \xmark & \xmark & \xmark & \cmark \\
R-4  & \cmark & \cmark & \cmark & \cmark \\
R-5  & \cmark & \xmark & \xmark & \cmark \\
R-6  & \cmark & \xmark & \cmark & \cmark \\
R-7  & \cmark & \xmark & \cmark & \cmark \\
R-8  & \cmark & \xmark & \xmark & \cmark \\
R-9  & \cmark & \xmark & \xmark & \cmark \\
R-10 & \xmark & \xmark & \xmark & \cmark \\
\bottomrule
\end{tabularx}
\end{table}

Table~\ref{tab:accuracy} reports the accuracy of LLM\textsubscript{2} under four feature vector extraction strategies. Specifically, \emph{One-stage One-inquiry} directly extracts VFVs from the law text using a single prompt, while \emph{One-stage Two-inquiry} first converts the law into a CVS and then extracts the feature vector within the same prompt. \emph{Two-stage One-inquiry} separates this process into two stages, where LLM\textsubscript{1} generates the CVS and LLM\textsubscript{2} extracts the VFV. Finally, \emph{Two-stage Multi-inquiry (Ours)} further decomposes the reasoning process into multiple fine-grained prompts, which are then aggregated to improve structured extraction.

The results show that the two-stage multi-inquiry strategy significantly outperforms the other approaches in conversion accuracy, confirming the effectiveness of first inferring semantic structure before performing structured extraction.

\begin{figure}[htbp]
    \centering
    \includegraphics[width=0.43\textwidth]{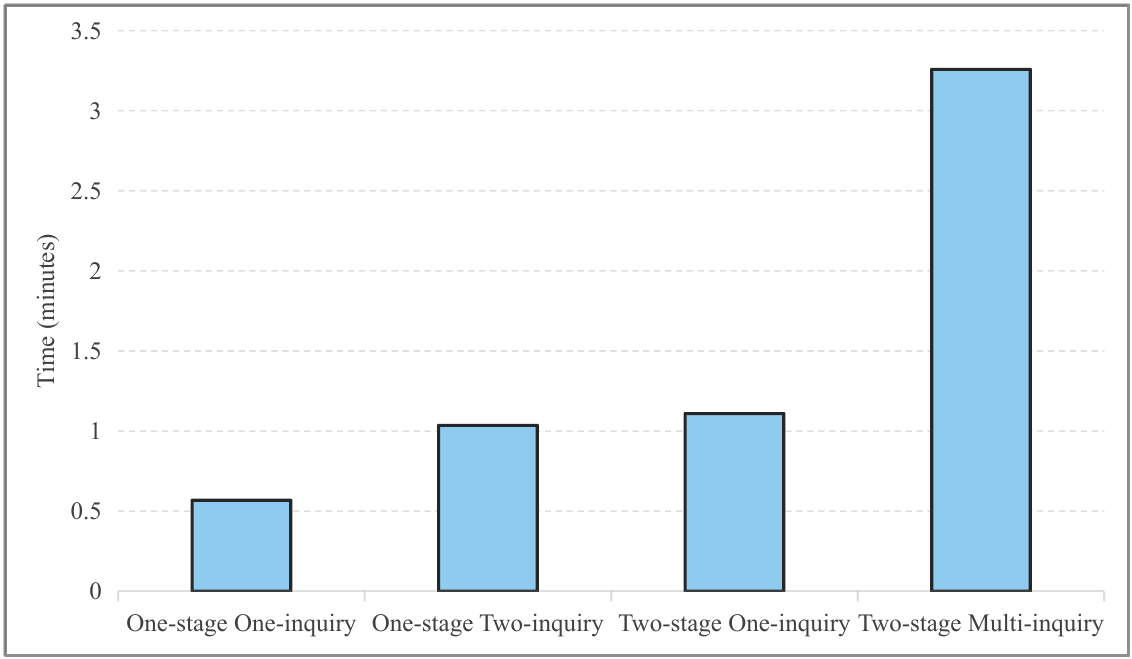}
    \caption{Average extraction time of different strategies}
    \label{fig:Average}
\end{figure}

Fig.~\ref{fig:Average} illustrates the average processing time for converting traffic laws into VFVs. While our method requires more inquiry steps, slightly increasing conversion time, this overhead is acceptable because it minimally affects overall generation efficiency while playing a critical role in ensuring high accuracy.

\section{Related Work}

\noindent\textbf{Large Language Models.}
LLMs have demonstrated strong capabilities in reasoning, code generation, and domain adaptation across diverse tasks~\cite{lemieux2023codamosa,liao2024dupllm,zuo2024exploring,arora2024optimizing,roziere2023code,yu2024fight}. Among adaptation techniques, prompt engineering~\cite{liu2023pre} offers a lightweight alternative to fine-tuning~\cite{radford2018improving}, as it steers model behavior without parameter updates~\cite{brown2020language}. In this work, we employ prompt engineering to leverage LLMs’ reasoning and generalization abilities for converting traffic laws into interpretable scenario representations.

\noindent\textbf{ADS Testing and Simulation.}
Existing ADS testing research mainly targets safety-critical failures such as collisions or near-miss events~\cite{gambi2019automatically,lu2022learning,chen2024misconfiguration,tang2021systematic,tang2021route,tang2023evoscenario}, often using real-world accident data to extract unsafe behaviors~\cite{tian2022generating,gambi2019generating,guo2024sovar}. However, such data-driven methods have limited coverage and generalizability. Law-based testing frameworks~\cite{zhang2023testing,sun2022lawbreaker,kim2022drivefuzz} attempt to incorporate regulatory semantics but rely on manual rules or deterministic inputs, reducing scalability. In contrast, our approach integrates traffic-law semantics with cooperative evolutionary fuzzing to systematically generate diverse violation scenarios.

\section{Threats to Validity}

This section discusses threats to validity of our evaluation, categorized as internal, construct, and external.

\noindent\textbf{Internal Validity.}
SLaFE leverages a two-stage LLM-based pipeline to extract VFVs from natural language traffic laws. Due to the inherent stochasticity of LLM inference (e.g., temperature-based sampling), outputs may vary slightly across runs. This introduces variability when scaling to large law sets. To mitigate this, we release all extracted vectors and law mappings to support replication and transparency.

\noindent\textbf{Construct Validity.}
SLaFE relies on LLM\textsubscript{1} to extract violation semantics, with initial evaluations conducted via user surveys. A second-stage validation is performed using LLM\textsubscript{2} with cross-checking by two authors. While this process reduces bias, it remains susceptible to human misinterpretation, especially for nuanced legal constraints. Although inter-annotator agreement is ensured, future work should incorporate expert review or multi-party cross-annotation to improve reliability.

\noindent\textbf{External Validity.}
Our evaluation is currently limited to the Apollo 7.0 stack within the LGSVL simulator. The generalizability of SLaFE to other autonomous driving platforms remains to be explored. Additionally, our region-semantic modeling is based on 10 representative traffic laws. Broader applicability requires further empirical validation across more diverse legal scenarios.

\section{Conclusion}
In this work, we presented SLaFE, a novel framework that integrates LLM-based semantic modeling with co-evolutionary fuzzing to uncover traffic law violations in ADS. SLaFE abstracts traffic laws into region-level semantics on San Francisco HD maps, where LLMs translate legal conditions into concrete triggering preconditions and behavioral constraints. These semantics are then encoded as feature vectors to enable cross-law co-evolution, effectively guiding AVs into potentially non-compliant driving scenarios. We evaluated SLaFE on ten real-world traffic regulations using the Apollo 7.0 platform. Experimental results demonstrate that SLaFE successfully induced all corresponding violations directly from natural language law descriptions, achieving a 100\% violation discovery rate.

\section*{Acknowledgment}

This work is supported by the National Natural Science Foundation of China under Grant No.62502550, No.62272145 and No.U21B2016, and by the Shenzhen Science and Technology Program (KJZD20240903095700001).

\bibliographystyle{IEEEtran}
\bibliography{sample-base}

\end{document}